\documentclass[apj,twocolumn]{emulateapj}
\usepackage{amsmath}
\usepackage{txfonts}

\begin{document}

\journalinfo{\sc
submitted to
\sl the Astrophysical Journal}

\shorttitle{\sc Central Velocity Dispersion}
\title{A Theorem on Central Velocity Dispersions}
\shortauthors{\sc J.~An \& N.W.~Evans}
\author{Jin~H.~An,\altaffilmark{1,2} and N.~Wyn~Evans\altaffilmark{3}}
\altaffiltext{1}{Dark Cosmology Centre, Niels Bohr Institute,
University of Copenhagen, Juliane Maries Vej 30, DK-2100 Copenhagen, Denmark}
\altaffiltext{2}{Niels Bohr International Academy, Niels Bohr Institute,
University of Copenhagen, Blegdamsvej 17, DK-2100 Copenhagen, Denmark}
\altaffiltext{3}{Institute of Astronomy, University of Cambridge,
Madingley Road, Cambridge, CB3~0HA, UK}

\slugcomment{\tt
jin@dark-cosmology.dk, nwe@ast.cam.ac.uk}

\begin{abstract}
It is shown that, if the tracer population is supported by
a spherical dark halo with a core or
a cusp diverging more slowly than that of a singular isothermal sphere,
the logarithmic cusp slope $\gamma$ of the tracers
must be given exactly by $\gamma=2\beta$ where $\beta$ is
their velocity anisotropy parameter at the center
unless the same tracers are dynamically cold at the center.
If the halo cusp diverges faster than that of
the singular isothermal sphere, the velocity dispersion of the tracers
must diverge at the center too. In particular, if the logarithmic halo
cusp slope is larger than two, the diverging velocity dispersion also
traces the behavior of the potential. The implication of our theorem
on projected quantities is also discussed. We argue that our theorem
should be understood as a warning against interpreting results based
on simplifying assumptions such as isotropy and spherical symmetry.
\end{abstract}


\section{Introduction}

Gravitationally-interacting collisionless N-body systems are a good
model for both dark-matter halos and stellar systems, and therefore
have been a subject of many studies. Early on, it was realized that
the evolution of such systems is governed by what is now known as the
collisionless Boltzmann equation \citep{Ed15,Je15}. However, since it
deals with the distribution function, which is generally inaccessible
in all but very extensive and nearly complete data sets, its utility
is somewhat limited in reality. An alternative to dealing directly
with the distribution function is to focus on the relations among the
statistical moments of it. The set of their governing equations is
obtained by taking velocity moment integrals on the collisionless
Boltzmann equation \citep[see e.g.,][]{BT} and bears the name of Sir
James H. Jeans (1877-1946). Of a particular interest among the set is
the one resulting from the first moment integral, which is what is
usually referred to as the Jeans equation. The equation summarizes the
momentum conservation in a local volume element and is an analogue to
the fluid/gas-dynamical Euler equation except for the presence of
anisotropic stress tensor relating to the local velocity dispersions
in place of the usual pressure term.

In recent years, interest in the Jeans equations has been rekindled for
various reasons. First, N-body simulations exhibit striking regularities
among dynamically relaxed dark matter structures
\citep[e.g.,][]{Ta01,HM06}. It is reasonable to suspect that these are
the results of some underlying physics working to erase the memory of
the initial conditions and settle the structure into a `universal'
form constrained by the Jeans equations \citep[e.g.,][]{De05,Ba06}.
Second, a wealth of new data has become available on the
velocities of giant stars in nearby dwarf spheroidals
\citep{Kl02,Wi04} and of planetary nebulae in nearby elliptical
galaxies \citep{Ro03,Dek05}. The structure of dark halos can be
mapped out through these data coupled with the Jeans equations. Of
course, it is of great interest to establish the dark halo structure in
the central regions, which is predicted to be cusped in hierarchical
cosmologies \citep[e.g.,][]{NFW,Mo98}.

All these have motivated greater theoretical scrutiny of the Jeans
equations. Recently, \citet{Ev08} have shown that any cusped profile
of an isothermal tracer population can be supported by the
potential generated by a dark halo only if the halo possesses an isothermal
cusp. In this paper, we extend their study and subject the properties of
systems governed by the spherical Jeans equations to a thorough
theoretical analysis.

\section{Jeans equations in spherical symmetry}

The Jeans equations under the spherical-symmetry and the steady-state
assumptions reduce to
\begin{equation}
\label{eq:je}
\frac{d(\nu\sigma_r^2)}{dr}
+2\beta\,\frac{\nu\sigma_r^2}r
=-\nu\frac{d\Psi}{dr}
\end{equation}
where $\nu=\nu(r)$ and $\sigma_r=\sigma_r(r)$ are the density profile
and the radial velocity dispersion of the tracer population. The
density here is nominally the number density, which follows the
derivation of the Jeans equations from the collisionless Boltzmann
equation. However, since equation (\ref{eq:je}) is linear in $\nu$, it
is still valid for any constant multiple of $\nu$, such as the mass or
the luminosity density, provided that the tracer population has
homogeneous properties.

The velocity anisotropy parameter $\beta=\beta(r)$ is defined such
that
\begin{equation}
\label{eq:be}
\beta=1-\frac{\sigma_\theta^2}{\sigma_r^2}
\end{equation}
where $\sigma_\theta(r)$ is the 1-d tangential velocity dispersion of
the same tracers \citep{BT}. The luminous tracers are moving in a
gravitational potential $\Psi(r)$, which, through the spherical
Poisson equation,
\begin{equation}
\label{eq:pe}
4\pi G\rho
=\frac1{r^2}\,\frac d{dr}\!\left\lgroup r^2\frac{d\Psi}{dr}\right\rgroup,
\end{equation}
is related to the density profile $\rho(r)$ of the dark halo.
Strictly speaking, $\rho$ in equation (\ref{eq:pe}) is the total mass
density that includes all gravitating masses. Here, it will be simply
referred to be a dark halo, which is basically a label that signifies
that we do \emph{not} demand that the potential should be
self-consistently generated by the density profile of the tracers.
Although physical solutions are still subject to constraints that
they must be non-negative and the tracer mass density may not be
greater than that of `dark halo' anywhere, these will not be
considered explicitly in this paper.

\section{Power-law solutions to Jeans equations}
\label{sec:gs}

We begin our study with an analysis of power-law solutions to the
Jeans equations with constant anisotropy parameter. These are often
good approximations locally, and have the advantage that they are
analytically tractable.

The spherical Jeans equation (\ref{eq:je}) is always formally
integrable such that
\begin{equation}
Q\nu\sigma_r^2=-\int\!dr\,Q\nu\frac{d\Psi}{dr}
\end{equation}
where $Q=Q(r)$ is the integrating factor
\begin{equation}
\ln Q=\int\!dr\,\frac{2\beta}r.
\end{equation}
If $\nu$ and $\Psi$ behave locally like a power law,
\begin{displaymath}
\nu\simeq Ar^{-\gamma}
\,;\qquad
\frac{d\Psi}{dr}\simeq\frac B{r^{\delta+1}}
\end{displaymath}
where $A,B>0$ and $\delta\le1$, we can find solutions to the Jeans
equations once the behavior for the anisotropy is prescribed. If the
potential is self-consistently generated, the power indices and
normalization constants are related to each other such that
$\gamma=\delta+2$ and $4\pi GA=B(1-\delta)$ (see eq.~[\ref{eq:pe}]).

The easiest assumption to make regarding the behavior of $\beta$ is
that it is constant. In reality, this probably is not true, but here
we are interested in the generic behavior of solutions for given local
power-law assumptions on $\nu$ and $\Psi$. Therefore, assuming
constant $\beta$ is valid provided that its variation is much slower
than that of the density and the potential. Notably, \citet{HM06} found
that the logarithmic density slope and the anisotropy parameter are
linearly related in simulated dark halos. Although their detailed
finding appears not to be always true \citep[see e.g.,][]{Na08}, the
general idea seems to be still valid. That is to say, the spatial
variation of the anisotropy parameter is sufficiently slow so that it
can be considered to be locally constant while the density profile is
approximated as a power law. The situation is believed not to be much
different in stellar systems, for which no evidence to the contrary is
obvious.

Under the assumption of the constancy of $\beta$, we have
$Q=r^{2\beta}$, which results in
\begin{equation}
\label{eq:gen_sol}
\sigma_r^2\simeq
\begin{cases}\
\displaystyle{\frac B{\gamma-2\beta+\delta}\,\frac1{r^\delta}
+Cr^{\gamma-2\beta}}
&\text{if $\delta\ne2\beta-\gamma$};
\smallskip\\
r^{\gamma-2\beta}\left(B\ln r^{-1}+C\right)
&\text{if $\delta=2\beta-\gamma$},
\end{cases}
\end{equation}
where $C$ is an integration constant to be determined from the boundary
condition.

While the solution in equation (\ref{eq:gen_sol}) is always valid for given
assumptions, it is somewhat easier to follow its behavior if we
consider a finitely-deep and an infinitely-deep potential well
separately. In the next two subsections, we investigate each case
in detail, and find the constraints on the behavior of the radial
velocity dispersion provided by the Jeans equations.

\subsection{Case 1: finite central potential wells}

If the dark halo diverges like a singular isothermal sphere (which
behaves as $r^{-2}$; henceforth SIS), the resulting potential is
logarithmically-divergent and so $\delta=0$. Thus, given the power-law
assumption, for any cusped halo diverging slower than a SIS, we can
limit $\delta<0$. Letting $p=-\delta>0$ (that is, the potential
behaves like $\Psi\simeq\Psi_0+Br^p/p$), the leading term for
$\sigma_r^2$ as $r\rightarrow0$ becomes
\begin{equation}
\sigma_r^2\approx
\begin{cases}\
\displaystyle{\frac B{\gamma-2\beta-p}\,r^p}
&\text{if $p<\gamma-2\beta$};
\smallskip\\
r^p\left(B\ln r^{-1}+C\right)
&\text{if $p=\gamma-2\beta$};
\smallskip\\
Cr^{\gamma-2\beta}
&\text{if $\gamma-2\beta<p$}.
\end{cases}
\end{equation}
For $0<p<\gamma-2\beta$, we have $\sigma_r^2\sim r^p\rightarrow0$ as
$r\rightarrow0$. If $p=\gamma-2\beta>0$, then $\sigma_r^2\sim r^p\ln r^{-1}$
and again $\lim_{r\rightarrow0}\sigma_r^2\rightarrow0$. Moreover,
the logarithmic slope of $\sigma_r^2$ still tends to $p>0$ in the
limit of $r\rightarrow0$. Finally, if $\gamma-2\beta<p$, then
$\sigma_r^2\sim r^{\gamma-2\beta}$. However, for this last case, if
$\gamma<2\beta$, then $r^{\gamma-2\beta}\rightarrow\infty$ as
$r\rightarrow0$. Since any finitely-deep central potential well is
unable to support tracer populations with divergent velocity
dispersions, the physical possibilities are limited to be
$\gamma\ge2\beta$ \citep[c.f,][]{AE06}.

In conclusion, with a finite central potential well, the possibilities
are either (i) $\sigma_r^2\rightarrow0$ as $r\rightarrow0$ [here, the
logarithmic slope of $\sigma_r^2$ tends to $\min(p,\gamma-2\beta)>0$]
or (ii) $\gamma=2\beta$ with finite and non-zero $\sigma_r^2$ at $r=0$
\citep[see also][]{Ev08}.

\subsection{Case 2: centrally divergent potentials}

Next, we consider a centrally-diverging potential, for which
$0\le\delta\le1$. Here, the $\delta=0$ case corresponds to a
logarithmic potential and a SIS-like dark halo cusp, whereas a point
mass potential is represented by $\delta=1$.

For these cases, the theorem of \citet{AE06} provides us with the
constraint that $\gamma\ge\case12\delta+\beta(2-\delta)$, and so that
$\delta-2\beta+\gamma\ge(\case32-\beta)\delta$. Since $\beta\le1$, we
finally find that $\delta-2\beta+\gamma\ge0$. Here, this is strictly
larger than zero if $\delta\ne0$. Consequently, the leading term for
$\sigma_r^2$ as $r\rightarrow0$ with a divergent potential is
\begin{equation}
\label{eq:lgp}
\sigma_r^2\simeq
\begin{cases}\
\displaystyle{\frac B{\gamma-2\beta}+Cr^{\gamma-2\beta}}
&\text{if $\delta=0$ and $\gamma>2\beta$};
\smallskip\\
B\ln r^{-1}+C
&\text{if $\delta=\gamma-2\beta=0$},
\end{cases}
\end{equation}
or
\begin{equation}
\sigma_r^2\approx
\frac B{\gamma-2\beta+\delta}\,\frac1{r^\delta}
=\frac\delta{\gamma-2\beta+\delta}\,|\Psi|
\end{equation}
if $0<\delta\le1$, for which $\gamma-2\beta+\delta>0$.
Here, note that $\Psi\simeq B\ln r^{-1}$ if $\delta=0$ and
$\Psi\simeq-(B/\delta)r^{-\delta}$ for $0<\delta\le1$. In addition,
$\rho\propto r^{-(2+\delta)}$ for $0\le\delta<1$ if the potential is
generated by $\rho$, whereas $B=GM_\bullet$ and $\delta=1$ if a black
hole of mass $M_\bullet$ dominates the potential.

In conclusion, $\sigma_r^2$ in a divergent potential well traces the
potential except for the case of a logarithmically-divergent potential
with $\gamma>2\beta$, for which it is non-zero and finite.

\section{Properties of solutions\\ to Jeans equations at center}

The preceding results are interesting, but the arguments leading to
them are restricted to power-law solutions of the Jeans equations with
constant anisotropy parameter. In fact, the results hold good more
generally albeit in a weaker form. In the following, we shall
derive a general theorem that extends the preceding results.
In this section, we will set up the framework and proceed to the case
concerning the halo with a core or a cusp that is shallower than the
SIS, which is likely to encompass most astrophysically interesting
models. We will extend the theorem to all physically allowed models of
halos and resulting potentials including those dominated by a central
point mass in \S~\ref{sec:gencase}. The theorem is deduced by
analyzing the Jeans equation (\ref{eq:je}) in the limit of
$r\rightarrow0$ without reference to constancy
of the anisotropy parameter or power-law behaviors. The resulting
constraints however are only strictly applicable to the central
limiting values.

\subsection{Preliminaries}

First, let us recast the Jeans equation into a more useful form. We
begin by integrating equation (\ref{eq:pe}), which leads us to
\begin{equation}
\label{eq:mass}
r^2\frac{d\Psi}{dr}=
GM(r)=GM_\bullet
+4\pi G\int_0^r\!d\tilde r\,\tilde r^2\rho(\tilde r)
\end{equation}
where $M(r)$ is the enclosed mass within the radius of $r$, and the
integration constant $M_\bullet$ represents a central point mass
(e.g., a supermassive black hole). However, we postpone detailed
consideration of the $M_\bullet\neq0$ case until \S~\ref{sec:gencase}.

Equation (\ref{eq:je}) is then equivalent, with the enclosed mass, to;
\begin{equation}
\label{eq:master}
\frac{GM(r)}r=\sigma_r^2
(\gamma-2\beta-\alpha).
\end{equation}
Also we have introduced the logarithmic slopes of the tracer
density and the velocity dispersion, namely (note the signs)
\begin{equation}
\label{eq:ga}
\gamma=-\frac{d\ln\nu}{d\ln r}
=-\frac r\nu\frac{d\nu}{dr}
\,;\qquad
\alpha=\frac{d\ln\sigma_r^2}{d\ln r}
=\frac r{\sigma_r^2}\frac{d\sigma_r^2}{dr}.
\end{equation}
%

In the following, we consider the behavior of the system at the
center, as indicated by the limit of equation (\ref{eq:master}) as
$r\rightarrow0$. All the subsequent arguments operate under the
assumption that every quantity considered here is well-behaved,
continuous and smooth.

\subsection{Systems with vanishing $M/r$ at the center}
\label{sec:cvd1}

Here, we basically repeat the argument found in the section~5 of
\citet{Ev08} with a slight refinement. The result will form a part of
the theorem to be proven in \S~\ref{sec:gencase}, and highlights the
astrophysically relevant information.

The condition for the left-hand side of equation (\ref{eq:master}) to
vanish in the limit $r\rightarrow0$ is given by
$\lim_{r\rightarrow0}M(r)=M_\bullet=0$ and, from l'H\^opital's rule,
$dM/dr|_{r=0}=0$. The last bit is equivalent to
$\lim_{r\rightarrow0}\rho r^2=0$ -- that is to say,
$\lim_{r\rightarrow0}\rho$ is finite (i.e., a cored profile) or $\rho$
diverges at the center slower than a SIS. Hence, assuming
$M_\bullet=0$ (i.e., no central point mass), if
$\lim_{r\rightarrow0}\rho r^2=0$, then the right-hand side of equation
(\ref{eq:master}) should also vanish as $r\rightarrow0$. This is
possible only if (i) $\sigma^2_{r,0}=0$, or (ii)
$\alpha_0=\gamma_0-2\beta_0$. Here and throughout, the subscript
``0'' is used to indicate the limiting value at the center.

Suppose that (ii) is the case. Here, if $\gamma_0<2\beta_0$, then it
would be that $\sigma_r^2\sim r^{-(2\beta_0-\gamma_0)}\rightarrow\infty$
as $r\rightarrow0$. However, the SIS, for which
$\lim_{r\rightarrow0}\rho r^2$ is non-zero and finite, can only
generate a potential diverging as fast as logarithmic. Thus, the
velocity dispersion that diverges like a power law cannot be supported
by the potential generated by the density cusp that is shallower than
that of the SIS. Consequently, the corresponding case,
$\gamma_0<2\beta_0$, is unphysical and not allowed. On the other hand,
if $\gamma_0>2\beta_0$, then $\sigma_r^2\sim
r^{(\gamma_0-2\beta_0)}\rightarrow0$, which reduces to the case (i).
In conclusion, a spherical dark halo with a core or a milder cusp than
that of a SIS (i.e., $\lim_{r\rightarrow0}\rho r^2=0$), can only permit
tracer populations satisfying the constraint $\beta_0=\gamma_0/2$ or
those with $\sigma^2_{r,0}=0$.

\subsection{Vanishing central velocity dispersions}

The preceding discussion indicates that $\sigma^2_{r,0}=0$ is
necessary for $\lim_{r\rightarrow0}\rho r^2=0$ if
$\gamma_0\ne2\beta_0$. The initial impression of equation
(\ref{eq:master}) notwithstanding, $\sigma^2_{r,0}=0$ alone however is
not sufficient for vanishing $M/r$, either. Formally, this is because
the behavior of equation (\ref{eq:master}) cannot be specified for
tracers for which $\sigma^2_{r,0}=0$ \emph{and} $\beta_0=-\infty$
without reference to the speed of each approach to its limiting value.

From its definition (eq.~[\ref{eq:be}]), $\beta=-\infty$ if
$\sigma_r^2=0$ unless $\sigma_\theta^2=0$ too. Here, $\beta_0=-\infty$
indicates a tracer population with purely circular orbits toward the
center. Note that in principle it is always possible to construct any
spherical model with purely circular orbits (although such models are
subject to resonant over-stabilities; see \citealt{Pa89}). The
non-zero tangential velocity dispersion here is the result of the
random orientations of the orbital planes while the circular speed is
uniquely specified by the enclosed mass. That is to say, the local
dispersion of the \emph{speed} of the tracers is actually zero
although the tangential velocity dispersion may be non-zero.

On the other hand, for all finite values of $\beta_0$, then matters
are simpler. Since
\begin{displaymath}
\sigma^2_{\theta,0}=\sigma^2_{\phi,0}=(1-\beta_0)\sigma^2_{r,0}=0,
\end{displaymath}
we easily find that $\sigma^2_{r,0}=0$ indicates that the total
3-d velocity dispersion at the center also vanishes.
That is, the system must be dynamically cold at the center.

The conclusion of \S~\ref{sec:cvd1} therefore may be rephrased as
follows: in a spherical potential well generated by a halo that is
cored or cusped less severe than a SIS, the only allowed populations
of tracers are those either consisting of purely circular orbits
toward the center or exhibiting vanishing central velocity
dispersions unless the limiting values of the cusp power index
$\gamma_0$ and the anisotropy parameter $\beta_0$ at the center are
constrained such that $\gamma_0=2\beta_0$.

\subsection{Discussion}

At first glance, the result of \S~\ref{sec:cvd1} may appear to be
counterintuitive as though it seems to suggest no pressure support at
the center whereas the Jeans equation is supposed to balance the
force. This reasoning is faulty because the actual `pressure' in this
case is given by $\rho\sigma^2$, not $\sigma^2$. Given the density
cusp, the system possesses non-vanishing kinematic pressure at the
center even though it is dynamically cold. This is obvious in the
power-law solutions to the Jeans equation of \S~\ref{sec:gs},
which show that $\nu\sigma_r^2\sim r^{\min(p-\gamma,-2\beta)}$. For a
system that is radially-biased or isotropic toward the center (i.e.,
$\beta\ge0$, for which $\gamma\ge0$ from \citealt{AE06} and so tracers
with a hole at the center are not allowed), the central `pressure' is
therefore always non-vanishing as $r\rightarrow0$ (it would be actually
divergent unless $\beta=0<p-\gamma$). The `pressure' of the
tangentially-biased system ($\beta<0$) on the other hand can be
vanishing if $p>\gamma$. However, in this last scenario, it can be
understood that the tangentially-biased system is preferentially
rotationally-supported toward the center.

If $\sigma_r^2$ indeed vanishes at the center, this implies that there
are no radial orbits in the model and that the distribution function
has the property $\lim_{L^2\rightarrow0}f(L^2,E)\rightarrow0$. This
seems unusual, but there are mechanisms known that can depopulate the
radial orbits -- for example, scattering by the central cusp
\citep{Ge85} or the radial orbit instability \citep{Pa87} -- albeit at
the cost of driving the system away from sphericity.

The result can also be applied to a self-consistent system (or
equivalently interpreted as a constraint on the central velocity
dispersion of the dark halo itself). In a spherical dark halo that has
a core or a cusp less severe than a SIS, the central limiting value of
the anisotropy parameter must be exactly half of the numerical value
of the cusp slope unless the central velocity dispersion vanishes. In
particular, a cored halo must have an isotropic velocity dispersion at
the center if it is not dynamically cold there.

\section{The general theorem}
\label{sec:gencase}

So far, we have investigated some astrophysically important
subcases. Here, we derive and prove the general theorem that makes
precise the interlocking constraints between the central limiting
values of the density and velocity dispersion of the tracers and the
potential. Those who are primarily interested in the result should
skip to \S~\ref{sec:the}. The next subsection provides a rigorous
mathematical analysis of equation (\ref{eq:master}) that leads to our
result.

\subsection{A derivation of the theorem}

We start by giving the binary relation `$\sim$' its precise
mathematical meaning. In the following, it is understood to be the
short-hand notation such that $a\sim b$ as $r\rightarrow0$ if and only
if both $\lim_{r\rightarrow0}(a/b)$ and $\lim_{r\rightarrow0}(b/a)$
are finite. We also define the binary relations $\succnsim$ and
$\precnsim$ similarly. That is to say, $a\succnsim b$ (or $a\precnsim
b$) as $r\rightarrow0$ if and only if $\lim_{r\rightarrow0}a/b$ is
divergent (or zero). In addition, it is also to be understood that the
limit is taken to be always $r\rightarrow0$ unless the explicit
reference to the limit is given to override.

Next, we consider the behavior of the left-hand side of equation
(\ref{eq:master}) in relation to the mass density profile that
generates the potential. From the argument of \S~\ref{sec:cvd1}, we
find that $M/r$ decays to zero in the limit $r\rightarrow0$ if
$M_\bullet=0$ and $\rho\precnsim r^{-2}$. The corresponding potential
is either finite at the center or diverges strictly slower than
logarithmic (i.e., $\Psi\precnsim\ln r$). On the other hand, it
attains a non-zero finite limiting value at $r\rightarrow0$ if and
only if $M(r)\sim r$. This is equivalent to $\rho\sim r^{-2}$ and also
$\Psi\sim\ln r$. Finally, $M/r$ diverges as $r\rightarrow0$ if
$M_\bullet$ is non-zero (implying the presence of a central point
mass) or $\rho\succnsim r^{-2}$ in the same limit. The corresponding
potential also diverges but strictly faster than logarithmic (i.e.,
$\Psi\succnsim\ln r$) in the same limit.

The behaviors of $M/r$ and $\Psi$ in relation to each other and their
respective logarithmic slopes may be explored in further detail. Let
us first define $p$, the logarithmic slope of $M/r$, i.e.,
\begin{equation}
\label{eq:defp}
p=\frac{d\ln(M/r)}{d\ln r}
=\frac1{d\Psi/dr}\,\frac d{dr}\!\left\lgroup\frac{GM}r\right\rgroup.
\end{equation}
Note that in the limit of $r\rightarrow0$, the logarithmic slope of
the potential (difference) necessarily tends to the same value as
$p_0$. In particular, if $\Psi(0)=\Psi_0$ is finite, it naturally
follows that $GM/r=r(d\Psi/dr)\rightarrow0$ and so
\begin{displaymath}
\lim_{r\rightarrow0}\frac{d\ln|\Psi-\Psi_0|}{d\ln r}
=\lim_{r\rightarrow0}\frac{GM/r}{\Psi-\Psi_0}=p_0\ge0
\end{displaymath}
using l'H\^opital's rule. If $\lim_{r\rightarrow0}\Psi=-\infty$ on the
other hand, we also find that
\begin{displaymath}
\lim_{r\rightarrow0}\frac{d\ln|\Psi|}{d\ln r}
=\lim_{r\rightarrow0}\frac{GM/r}\Psi=p_0\le0.
\end{displaymath}
Although the usual conditions for l'H\^opital's rule for this case are
only strictly met if $M/r$ diverges, l'H\^opital's rule can in fact be
proven only assuming a divergent denominator. Thus, the result holds
even though $M/r$ tends to a finite value (including zero) as
$r\rightarrow0$. Moreover, it is obvious that $p_0=0$ if
$\Psi\rightarrow\infty$ and $M/r$ is finite as $r\rightarrow0$. Next,
if $p_0\ne0$, then $M/r\sim|\Delta\Psi|$ where $\Delta\Psi=\Psi$ for
$\lim_{r\rightarrow0}\Psi=-\infty$ or $\Delta\Psi=|\Psi-\Psi_0|$ for
$\Psi(0)=\Psi_0$ being finite. By contrast, that $p_0=0$ indicates
that $M/r\precnsim|\Delta\Psi|$. If the central potential is
additionally finite (e.g, $\Psi-\Psi_0\sim[\ln r^{-1}]^{-1}$, for
which $M/r\sim[\ln r]^{-2}$), then $M/r\rightarrow0$. The behavior of
$M/r$ as $r\rightarrow0$ for a divergent central potential depend on
how fast $\Psi$ diverges relative to logarithmic divergence ($\ln
r^{-1}$) -- e.g., for $|\Psi|\sim\ln\ln r^{-1}$, $|\Psi|\sim\ln r$,
and $|\Psi|\sim\case12[\ln r]^2$, we have that $M/r\sim[\ln
r^{-1}]^{-1}$, $M\sim r$ and $M/r\sim\ln r^{-1}$, respectively.

Next, we proceed to analyzing the right-hand side of equation
(\ref{eq:master}). Here, we do not consider the $\beta=-\infty$ case,
which represents the formal possibility of building the system with
purely circular orbits. Then, since $M$ and $\sigma_r^2$ must be
non-negative, $\gamma_0-2\beta_0-\alpha_0\ge0$. Moreover,
$\gamma_0\le3$ from the constraint that the central mass concentration
must be finite. Together with the assumption that $\beta_0$ is finite,
we find that $(\gamma_0-2\beta_0-\alpha_0)$ is divergent only if
$\alpha_0$ diverges to negative infinity. However, then
$\sigma_r^2$ diverges faster than any power law to an essential
singularity (e.g., $e^{1/r}$). This is physically impossible, because
no real potential diverges faster than $1/r$ nor is thus able to
support such steeply diverging velocity dispersions. Therefore, we
limit $\gamma_0-2\beta_0-\alpha_0$ to be finite.

If $\gamma_0-2\beta_0-\alpha_0>0$, it is clear that $\sigma_r^2\sim
GM/r$, which also indicates that $\alpha_0=p_0$. If
$\alpha_0=p_0\ne0$, then $\sigma_r^2\sim|\Delta\Psi|$, too. If
$\alpha_0=p_0=0$ on the other hand, the behavior of $\sigma_r^2$ still
traces that of $M/r$, but $\sigma_r^2\sim M/r\precnsim|\Delta\Psi|$.

If $\gamma_0-2\beta_0=\alpha_0$, then $\sigma_r^2\succnsim M/r$ and so
$\alpha_0\le p_0$. Here, if the central potential is finite, then we
find that $0\le\alpha_0=\gamma_0-2\beta_0\le p_0$, from the constraint
of \citet{AE06}. For a divergent potential (introducing
$\delta_0=-p_0$, for which $0\le\delta_0\le1$), the constraint of
\citet{AE06}, $\gamma_0\ge\case12\delta_0+\beta_0(2-\delta_0)$,
indicates that $\case12\delta_0-\beta_0\delta_0\le\gamma_0-2\beta_0=
\alpha_0\le p_0=-\delta_0$. Now, if $\delta_0>0$, this would imply
$\beta_0\ge\case32$. This is obviously impossible, and therefore
$\delta_0=p_0=\alpha_0=\gamma_0-2\beta_0=0$. In addition, if
$M\succnsim r$, it is clear that $\sigma_r^2\rightarrow\infty$.
Furthermore, from equation (\ref{eq:master}) recast to be
\begin{displaymath}
r\frac{d\Psi}{dr}=(\gamma-2\beta)\sigma_r^2
-r\,\frac{d\sigma_r^2}{dr},
\end{displaymath}
we find for sufficiently-fast-decaying $\gamma-2\beta$ that
$\sigma_r^2\sim|\Psi|$. This essentially implies that $\sigma_r^2$
cannot diverge faster than $\Psi$.

\subsection{The statement of the theorem}
\label{sec:the}

In summary,
%
the spherical Jeans equations permit only restricted physical
possibilities regarding the limiting behaviors at the center. In
particular, the central limiting value of the velocity anisotropy
($\beta_0$; eq.~[\ref{eq:be}]) and those of the logarithmic slopes
of the luminous tracer density ($\gamma_0$; eq.~[\ref{eq:ga}]), the
radial velocity dispersion ($\alpha_0$; eq.~[\ref{eq:ga}]) and the
potential ($p_0$; eq.~[\ref{eq:defp}]) must meet
one, and only one, of the following
list of choices,
\begin{enumerate}
\renewcommand{\labelenumi}{\theenumi}
\renewcommand{\theenumi}{(\roman{enumi})}
\item $p_0=\alpha_0<\gamma_0-2\beta_0$ and
$\sigma_r^2\sim M/r$,
\item $p_0\ge\alpha_0=\gamma_0-2\beta_0\ge0$ and
$\Psi_0$ is finite,
\item $p_0=\alpha_0=\gamma_0-2\beta_0=0$ and
$\lim_{r\rightarrow0}\Psi=-\infty$,
\item $\beta_0=-\infty$.
\end{enumerate}
%
Focusing on the behavior of the velocity dispersion, the result with
the proviso $\beta>-\infty$ is summarized as
\begin{equation}
\lim_{r\rightarrow0}\frac{d\ln\sigma_r^2}{d\ln r}=
\begin{cases}\
\min(2-\Gamma_0,\gamma_0-2\beta_0)\ge0
&(\Gamma_0<2)\\
-(\Gamma_0-2)\le\gamma_0-2\beta_0
&(\Gamma_0\ge2)
\end{cases}
\end{equation}
although this does not include all the information encompassed in the
above choices. Here,
\begin{displaymath}
\Gamma=-\frac{d\ln\rho}{d\ln r}
\end{displaymath}
so that $p=2-\Gamma$, and extending to include the central point mass
by setting $\Gamma_0=3$.

For a prescribed behavior of $M/r$ or $\Psi$, the above list returns
the natural extension and generalization of our earlier results. If
$M/r\rightarrow0$ for example, then either 1) $\sigma_r^2\rightarrow0$
with $\sigma_r^2\sim M/r$ or $\alpha_0=\gamma_0-2\beta_0>0$, or 2)
$\alpha_0=\gamma_0-2\beta_0=0$. Consequently, we recover the
conclusion of \S~\ref{sec:cvd1}. The implication of the list however
is more detailed. First, if $\gamma_0-2\beta_0>p_0\ge0$ ($p_0\ge0$ is
necessary for vanishing $M/r$), then only the case (i) is possible and
so $\sigma_r^2\sim M/r$ and $\alpha_0=p_0$. If $p_0>0$ additionally,
then $\sigma_r^2\sim|\Psi-\Psi_0|$ with a finite $\Psi_0$ whereas
$\sigma_r^2\precnsim|\Delta\Psi|$ for $p_0=0$. On the other hand, with
$p_0\ge\gamma_0-2\beta_0>0$, we have $\alpha_0=\gamma_0-2\beta_0>0$
and so $\sigma_r^2\rightarrow0$ (and $p_0>0$ similarly indicating that
$M/r\sim|\Psi-\Psi_0|$). The remaining physical possibility,
$p_0\ge\gamma_0-2\beta_0=0$, implies that $\alpha_0=0$, which can lead
to a non-zero finite limit for $\lim_{r\rightarrow0}\sigma_r^2$. If
$\Psi_0$ is finite, then $\sigma_r^2$ must not diverge, but if $\Psi$
is divergent (but not faster than logarithmic) as $r\rightarrow0$, the
exact limiting behavior of the corresponding $\sigma_r^2$ should be
inferred from the particular solution to the Jeans equations.

By contrast, if $M/r$ diverges (for which $\Psi\rightarrow-\infty$),
then case (i) indicates that $\sigma_r^2\sim M/r\rightarrow\infty$
whereas case (iii) requires $\sigma_r^2\succnsim M/r$ and so
$\sigma_r^2\rightarrow\infty$ (but $\alpha_0=p_0=0$). In other words,
$\sigma_r^2$ necessarily diverges if $M/r\rightarrow\infty$.
Furthermore, $\sigma_r^2$ must be divergent as fast as $M/r$ (note
that if $\delta_0>0$, then $\sigma_r^2\sim M/r\sim|\Psi|$, but
$\sigma_r^2\sim M/r\precnsim|\Psi|$ for $\delta_0=0$ where $\delta_0$
is the negative logarithmic slope of $M/r$ or the potential) unless
$\delta_0=\gamma_0-2\beta_0=0$ for which $\sigma_r^2$ diverges faster
than $M/r$ but not faster than $|\Psi|$.

For $M\sim r$ (and $\Psi\sim\ln r$), the result is basically that of
equation (\ref{eq:lgp}); case (i) yielding the possibility of a finite
limiting value of $\sigma_r^2$ whereas case (iii) is consistent with
$\sigma_r^2$ diverging at most logarithmically or slower.

\subsection{Infinite velocity dispersions?}

In the framework of classical Newtonian mechanics upon which the Jeans
equations and the collisionless Boltzmann equation are ultimately
based, the divergence of $\sigma_r^2$ when $M/r$ and the corresponding
potential also diverge is in principle physically acceptable despite
its mathematical quirk. However, it is clear that the arguments given
in this paper eventually break down as $\sigma_r$ approaches the
speed of light. Moreover, in the corresponding halo,
$M/r$ should be divergent as $r\rightarrow0$, and therefore
there exists a radius below which $GM(r)/c^2>r$. Consequently,
the central cusp, if it ever were present, must collapse to a
singularity. In other words, one would expect that the formal infinity
of the velocity dispersion can be always circumvented through the
presence of a central black hole. The proper examination of physical
behaviors of the tracers and the halo under these conditions would
require consideration of relativistic physics, which is out of the
scope of the current paper. Of course, in reality, it is more likely
that other various physical complexities in the system intervene to
prevent the spherical Jeans equations to be applied uncritically all
the way down to the center even before any relativistic effects
become important.

\section{Projected quantities}

The direct measurement of radial and tangential velocity dispersions
of stellar tracers is limited to nearby populations. More generally,
the true observables are limited to the line-of-sight velocity
dispersion -- either the `aperture-averaged' value or its profile for
a subset. The implication of the theorem on the behavior of the
line-of-sight velocity dispersion is therefore of a great practical
interest. However, we shall see that the integral transformation
involved in the line-of-sight velocity dispersion weakens the
theorem's practical constraints.

It is usually assumed that the observed line-of-sight velocity
dispersion follows the luminosity-weighted integration of the velocity
dispersions along the line-of-sight direction. The latter
$\sigma_\ell$ is mathematically well-defined quantity such that
\begin{gather}\label{eq:los}
\sigma_\ell^2(R)=\frac2I
\int_R^\infty\!\left(1-\beta\frac{R^2}{r^2}\right)\,
\frac{\nu\sigma_r^2r\,dr}{\sqrt{r^2-R^2}}
\\\intertext{where}\nonumber
I(R)=2\int_R^\infty\!\frac{\nu r\,dr}{\sqrt{r^2-R^2}}
\end{gather}
is the surface density of the tracers. If $\beta\sigma_r^2$ is
non-divergent, the leading term of $I\sigma_\ell^2$ as $R\rightarrow0$
cannot be dominant over that of the surface density $I(R)$.
Consequently, the leading term of $\sigma_\ell^2$ in the central limit
is largely dictated by the tracer density profile.

In particular, if the density profiles of the tracers are approximated
as power-law-like, we find the behavior of the leading terms for the
surface density to be \citep[see e.g.,][]{AZ09}
\begin{align*}
\nu&\simeq Ar^{-\gamma}&\rightarrow\
&\begin{cases}\
I\sim R^{-(\gamma-1)}&(\gamma>1)\\
I\sim\ln R^{-1}&(\gamma=1)\\
I\sim I_0-I_1R^{1-\gamma}&(0<\gamma<1)
\end{cases}\\
\nu&\simeq\nu_0-Ar^q&\rightarrow\
&\begin{cases}\
I\sim I_0-I_1R^{1+q}&(0<q<1)\\
I\sim I_0-I_1R^2\ln R^{-1}&(q=1)\\
I\sim I_0-I_1R^2&(q>1)
\end{cases}
\end{align*}
where $I_0$ is the finite central surface density, and $A$ and $I_1$
are some positive constants. Assuming $\nu\sim r^{-\gamma}$
($\gamma=0$ if cored), $\sigma_r^2\sim r^\alpha$ ($\alpha>0$), and
$\beta_0>-\infty$, the corresponding behavior for equation (\ref{eq:los})
is similarly found to be
\begin{equation}
I\sigma_\ell^2\sim\begin{cases}\
R^{-(\gamma-\alpha-1)}&(\gamma>\alpha+1)\\
\ln R^{-1}&(\gamma=\alpha+1)\\
C_0+C_1R^{\alpha+1-\gamma}&(\alpha-1<\gamma<\alpha+1)\\
C_0+C_1R^2\ln R^{-1}&(\gamma=\alpha-1)\\
C_0+C_1R^2&(\gamma<\alpha-1)
\end{cases}
\end{equation}
with $C_0$ and $C_1$ being some non-zero constants. Given the
implication of the theorem for tracers with $\gamma_0\ne2\beta_0$ in a
non-divergent potential, i.e., $\sigma_r\rightarrow0$ as
$r\rightarrow0$ and so $\alpha>0$, we surmise that
$\sigma_\ell\rightarrow0$ ($\sim R^{\min(\gamma-1,\alpha)}$) as
$R\rightarrow0$ if $\gamma\ge1$ whereas it attains a finite limiting
value (and typically increasing outward) if $\gamma<1$. If on the
other hand $\sigma_{r,0}^2$ is finite (for which $\gamma_0=2\beta_0$
according to the theorem) or $\beta_0=-\infty$ (and
$\sigma_{\theta,0}^2=\sigma_{\phi,0}^2$ is non-zero), the leading term
behavior of $I\sigma_\ell^2$ is similar to that of $I$ alone, and thus
$\sigma_{\ell,0}^2$ is finite.

These essentially imply that the behavior of $\sigma_r^2$ cannot in
general be directly inferred from the leading term approximation of
$\sigma_\ell^2$ alone, and that the strict constraint from the theorem
is somewhat lost by going through the integral transformation.
Although one may deal with $I\sigma_\ell^2$ instead of $\sigma_\ell^2$
or can in principle invert the integral equation for $\sigma_\ell^2$
to yield $\sigma_r^2$ (assuming some particular $\beta$), this still
indicates that inferring $\sigma_r^2$ from $\sigma_\ell^2$ involves
analyzing higher-order behaviors of the latter and thus requires
high-precision measurements. Furthermore, this is independent of the
well-known degeneracy of $\sigma_r^2$ and $\beta$ in the inversion of
$\sigma_\ell^2$ in a sense that even though one possesses perfect a
priori information on $\beta$, the uncertainties in the recovered
$\sigma_r^2$ are always amplified by inverting $\sigma_\ell^2$.

\subsection{A central black hole}

The preceding discussion presumes the finite central potential well,
which is appropriate for the potential dominated by the halo that is
cored or cusped not so steep as the SIS. If the potential however is
dominated by the central point mass, the theorem indicates that
$\sigma_r^2\sim|\Psi|\sim 1/r$ and so $I\sigma_\ell^2\sim R^{-\gamma}$
for $\gamma>0$ (i.e., cusped tracer populations) or
$I\sigma_\ell^2\sim\ln R^{-1}$ for $\gamma=0$ (i.e., cored tracer
populations). That is to say, the line-of-sight velocity dispersion of
the population tracing the Keplerian potential is necessarily
divergent with its logarithmic slope, $|d\ln\sigma_\ell^2/d\ln R|$ being
equal to $\min(1,\gamma)$ where $\gamma$ is the 3-d density (negative)
logarithmic slope of the same tracers unless the orbits of tracers are
completely circularized toward the center. Nevertheless, the direct
application of this inference to the observational results warrants
caution since an assumption of `infinite' resolution is implicit in
the argument. That is to say, the result is strictly relevant only if
the observation can resolve the so-called sphere of influence of the
central point mass.

\section{Conclusions}

In this paper, we have established a general theorem -- stated in
\S~\ref{sec:the} -- that makes precise the relationship between the
central limiting values of the density and velocity dispersions of a
stellar population, together with the potential. Our theorem gives all
the mutually exclusive possibilities that can occur in a stellar
system. We note that our theorem has straightforward applications to
a number of astrophysical problems, including the kinematical modeling
of the stellar populations in dwarf spheroidal galaxies and elliptical
galaxies.

In \citet{Ev08}, we presented a simplified version of the theorem and
argued that it is the consequence of the spherical symmetry
assumption. However, after the appearance of the preprint version of
\citet{Ev08}, Scott Tremaine (priv.\ comm.) convinced us that the
theorem is due to the `non-analytic' point at the center. The
spherical symmetry is of secondary importance and only indirectly
responsible for the theorem by requiring a coordinate singularity at
the center. The theorem in this respect might be understood as an
incomplete boundary condition imposed on the Jeans equations at the
center resulting from the consideration of one-sided regularity.

Applying to the real astrophysical problems, the true moral of our
theorem is the urging of caution against interpreting results based on
simplifying assumptions. For instance, if one were to reconstruct the
dark halo density from the observations of the surface density and the
line-of-sight velocity dispersion profile of a tracer population, the
seemingly benign assumptions of spherical symmetry and isotropy
combined with a cored luminosity profile already severely restrict the
possible halo density (it cannot be cusped!). Such idealized
reconstructions are limited by the straitjacket imposed by the
theorem, yet the restrictions may be non-existent in reality -- not
unlike assuming a spherical cow!

\acknowledgments 
The authors thank S. Tremaine for comments made on \citet{Ev08}, which
led to the current work. The Dark Cosmology Centre is funded by the
Danish National Research Foundation.


\begin{thebibliography}{}

\bibitem[An \& Evans(2006)]{AE06}
An, J. H., \& Evans, N. W. 2006, \apj, 642, 752

\bibitem[An \& Zhao(2009)]{AZ09}
An, J. H., \& Zhao, H.-S. 2009, in prep.

\bibitem[Barnes et al.(2006)]{Ba06}
Barnes, E. I., et al.
2006, \apj, 643, 797

\bibitem[Binney \& Tremaine(2008)]{BT}
Binney, J., \& Tremaine, S. 2008, Galactic Dynamics, 2nd ed.\
(Princeton: Princeton Univ.\ Press)

\bibitem[Dehnen \& McLaughlin(2005)]{De05}
Dehnen, W., \& McLaughlin, D. E. 2005, \mnras, 363, 1057

\bibitem[Dekel et al.(2005)]{Dek05}
Dekel, A., et al.
2005, \nat, 437, 707

\bibitem[Eddington(1915)]{Ed15}
Eddington, A. S. 1915, \mnras, 76, 37

\bibitem[Evans et al.(2009)]{Ev08}
Evans, N. W., An, J., \& Walker, M. G. 2009, \mnras, 393, L50

\bibitem[Gerhard \& Binney(1985)]{Ge85}
Gerhard, O. E., \& Binney, J. 1985, \mnras, 216, 467

\bibitem[Hansen \& Moore(2006)]{HM06}
Hansen, S. H., \& Moore, B. 2006, New Astron., 11, 333

\bibitem[Jeans(1915)]{Je15}
Jeans, J. H. 1915, \mnras, 76, 70

\bibitem[Kleyna et al.(2002)]{Kl02}
Kleyna, J., et al.
2002, \mnras, 330, 792

\bibitem[Moore et al.(1998)]{Mo98}
Moore, B., et al.
1998, \apj, 499, L5

\bibitem[Navarro et al.(1995)]{NFW}
Navarro, J. F., Frenk, C. S., \& White, S. D. M. 1995, \mnras, 275, 720

\bibitem[Navarro et al.(2008)]{Na08}
Navarro, J. F., et al.
2008, \mnras, submitted
(arXiv:0810.1522)

\bibitem[Palmer \& Papaloizou(1987)]{Pa87}
Palmer, P. L., \& Papaloizou, J. 1987, \mnras, 224, 1043

\bibitem[Palmer et al.(1989)]{Pa89}
Palmer, P. L., Papaloizou, J., \& Allen, A. J. 1989, \mnras, 238, 1281

\bibitem[Romanowsky et al.(2003)]{Ro03}
Romanowsky, A. J., et al.
2003, Science, 301, 1696

\bibitem[Taylor \& Navarro(2001)]{Ta01}
Taylor, J. E., \& Navarro, J. F. 2001, \apj, 563, 483

\bibitem[Wilkinson et al.(2004)]{Wi04}
Wilkinson, M. I., et al.
2004, \apjl, 611, L21

\end{thebibliography}
\end{document}